\documentstyle[12pt]{article}
\begin{document}
\begin{center}
{\large \bf Theoretical issues of  small $x$ physics}
 \footnote{An introductory review talk given at the UK Phenomenology Workshop 
on HERA Physics, St. John's College, Durham, UK, September 1995}\\
\bigskip
J. Kwieci\'nski \\
Department of Physics\\ 
University of Durham\\
Durham, UK\\
and\\
Department of Theoretical Physics\\
H. Niewodnicza\'nski Institute of Nuclear Physics\\
Krak\'ow, Poland. \footnote{Permanent address}\\
\end{center}
\vspace{40mm}
\begin{abstract}
The perturbative QCD predictions concerning  deep inelastic scattering 
at low $x$ are summarized.   The theoretical framework based 
on the leading log $1/x$ resummation and $k_t$ factorization theorem 
is described and some recent developments concerning the BFKL equation 
and its generalization are discussed. The QCD expectations concerning the small 
$x$ behaviour of the spin dependent structure function $g_1(x,Q^2)$ are 
briefly summarized and the importance of the double logarithmic terms 
which sum contributions containing the leading powers of 
$\alpha_s ln^2(1/x)$ is emphasised. 
The role of studying final states in deep inelastic 
scattering for revealing the details of the underlying dynamics at low 
$x$ is pointed out  and some dedicated measurements, like deep inelastic 
scattering accompanied by an energetic jet, the measurement of the transverse 
energy flow etc.,   are briefly 
discussed.   
\end{abstract}
\newpage

  Perturbative QCD 
predicts that several new phenomena will occur when the parameter 
$x$ specifying 
the longitudinal momentum fraction of a hadron carried by a parton 
(i.e. by a quark or by a gluon) 
becomes very small \cite{GLR,BCKK} . The main expectation is
  that the gluon  densities 
should  strongly grow in this limit, eventually leading to the parton 
saturation effects \cite{GLR,BCKK,ADM1,JK1}. 
  The small $x$ behaviour of the structure functions 
is driven by the gluon through the $g \rightarrow q \bar q$ transition  
and the  increase of gluon distributions with decreasing $x$  
implies a similar increase of the deep inelastic lepton - proton  
 scattering structure function $F_2$ 
as the  Bjorken parameter $x$ decreases \cite{AKMS}.  
The Bjorken parameter $x$ 
is, as usual, defined  as $x=Q^2/(2pq)$ where $p$ is the proton four momentum, 
$q$ 
the four momentum transfer between the leptons and $Q^2=-q^2$.
The recent experimental data are consistent with 
this perturbative QCD prediction that the structure function $F_2(x,Q^2)$ 
should strongly grow with the decreasing Bjorken parameter 
$x$ \cite{H1,ZEUS}.\\      

 The growth of structure functions with decreasing parameter $x$ is 
much stronger than that 
which would follow from the expectations based on the "soft" pomeron 
exchange mechanism with the soft pomeron intercept 
$\alpha_{soft} \approx 1.08$ 
as determined from the phenomenological analysis of total hadronic 
and real photoproduction cross-sections \cite{DOLA}.\\

Small $x$ behaviour of structure functions  
for fixed $Q^2$ reflects the high energy behaviour of the 
virtual Compton scattering total cross-section with increasing   
total CM energy 
squared $W^2$ since $W^2=Q^2(1/x-1)$. The Regge pole exchange picture  
\cite{PC} would  
therefore appear quite appropriate for the theoretical description 
of this behaviour.\\

The high energy behaviour of the total 
hadronic and (real) photoproduction cross-sections can be economically 
described by two contributions: an (effective) pomeron with its 
intercept slightly above unity ($\sim 1.08$) and the leading meson 
Regge trajectories  
with  intercept $\alpha_R(0) \approx 0.5$ \cite{DOLA}.  
The reggeons can be 
identified as corresponding to $\rho, \omega$, $f$ or $A_2$ 
exchange(s) depending 
upon the quantum numbers involved. All these 
reggeons have approximately the same intercept.   
One refers to the pomeron obtained 
from the phenomenological analysis of  hadronic  total cross 
sections as the "soft" pomeron since the bulk of the processes building-up 
the cross sections are  low $p_t$ (soft) processes.\\

The Regge pole model gives the following parametrization of the deep 
inelastic scattering structure function $F_2(x,Q^2)$ at  
small $x$ 
\begin{equation}
F_2(x,Q^2)=\sum_i \tilde \beta_i(Q^2) x^{1-\alpha_i(0)}. 
\label{reggef}
\end{equation}
The relevant reggeons are 
those which can couple to two (virtual) photons.  The (singlet) part 
of the structure function $F_2$ is controlled at small $x$ by 
pomeron exchange, while the non-singlet part $F_2^{NS}=
F_2^p-F_2^n$ by the  $A_2$ reggeon.  
Neither pomeron nor $A_2$ reggeons  couple to the 
spin structure function $g_1(x,Q^2)$ which is described at small 
$x$ by the  exchange of reggeons corresponding to 
axial vector mesons \cite{IOFFE,EKARL} i.e. to  $A_1$ exchange for the non-singlet 
part $g_1^{NS} = g_1^{p}-g_1^{n}$ etc.
\begin{equation}
g_1^{NS}(x,Q^2)=\gamma(Q^2) x^{-\alpha_{A_1}(0)}. 
\label{gnsa1}
\end{equation}
  The  reggeons which correspond to axial vector mesons 
are expected to 
have very low intercept (i.e. $\alpha_{A_1} \le 0$ etc.).\\

The high energy behaviour which follows 
from perturbative QCD is often refered to as being related to the "hard" 
pomeron in contrast to the soft pomeron describing the high energy 
behaviour of hadronic and photoproduction cross-sections.\\

In Fig. 1 we summarize the present experimental situation on the 
variation of the total virtual Compton scattering cross-section  
with total CM  energy $W$ for different photon virtualities $Q^2$  which range 
from the real photoproduction ($Q^2=0$) to the deep inelastic region \cite
{ALEVY}. 
The change of high energy behaviour with the scale $Q^2$ is evidently 
present in the data.\\

The purpose of this talk is to review some of the  recent theoretical 
developments concerning  deep inelastic scattering in the limit of small 
$x$.  
The next section  is devoted to the discussion of the  
BFKL equation which resums the 
leading powers 
of $ln(1/x)$ and to  the unified evolution 
equation \cite{CIAF,CCFM,KMS1,BO} based  on coherence and angular ordering 
\cite{MTK}. The 
$k_t$ factorization theorem \cite{KTFAC,CIAFKT,MK}, its connection with collinear factorization as well 
as the small $x$ resummation effects within the QCD evolution formalism 
are also discussed in this section.   In sec. 3 we present 
the theoretical expectation for the small $x$ 
behaviour of the spin dependent structure function $g_1(x,Q^2)$ 
concentrating for  
simplicity on its non-singlet part.  The novel 
feature in this case is importance of the double logarithmic terms i.e. 
of those terms in the perturbative expansion which contain powers 
of $\alpha_sln^2(1/x)$.  Finally in sec. 4  we discuss some of the dedicated 
measurements like deep inelastic + jet events and transverse energy flow 
etc. which aim at revealing the dynamical details of  small $x$ physics.  
Sec. 5 contains a brief summary and outlook.  \\

\section*{2. The BFKL pomeron and QCD predictions for the small $x$ 
behaviour of the structure function $F_2$}
At small $x$ the dominant role is played by the gluons 
and the  basic dynamical quantity is the   
unintegrated gluon distribution 
$f(x,Q_t^2)$ where $x$ denotes the momentum fraction 
of a parent hadron carried by a gluon and $Q_t$  its transverse 
momentum.  The unintegrated distribution $f(x,Q_t^2)$ 
is related in the following way to the more familiar scale dependent 
gluon distribution $g(x,Q^2)$: 
\begin{equation}
xg(x,Q^2)=\int^{Q^2} {dQ_t^2\over Q_t^2} f(x,Q_t^2). 
\label{intg}
\end{equation}
In the leading $ln(1/x)$  approximation the unintegrated 
distribution $f(x,Q_t^2)$ satisfies 
the BFKL equation \cite{BFKL,LIPATOV,CIAF} which has the following form: 
$$
f(x,Q_t^2)=f^0(x,Q_t^2)+
$$
\begin{equation}
\bar \alpha_s \int_x^1{dx^{\prime}\over 
x^{\prime}} \int {d^2 q\over \pi q^2}
\left[{Q_t^2 \over (\mbox{\boldmath $q$}+
\mbox{\boldmath $Q_t$})^2} 
f(x^{\prime},(\mbox{\boldmath $q$}+
\mbox{\boldmath $Q_t$})^2)-f(x^{\prime},Q_t^2)\Theta(Q_t^2-q^2)\right]
\label{bfkl}
\end{equation}
where 
\begin{equation}
\bar \alpha_s={3\alpha_s\over \pi}
\label{alphab}
\end{equation}
The first and the second 
terms  on the right hand side of  eq. (\ref{bfkl}) correspond 
to  real gluon emission with $q$ being the transverse 
momentum of the emitted gluon, and to the virtual corrections respectively. 
$f^0(x,Q_t^2)$ is a suitably defined inhomogeneous term.\\

After resumming the virtual corrections and "unresolvable"  gluon 
emissions ($q^2 < \mu^2$)  where $\mu$ is the resolution 
defining the "resolvable" radiation,  equation (\ref {bfkl}) 
can be rearranged into the following "folded" form: 
$$
f(x,Q_t^2)=\hat f^0(x,Q_t^2)+
$$
\begin{equation}
 \bar \alpha_s
\int_x^1{dx^{\prime}\over 
x^{\prime}} \int {d^2 q\over \pi q^2} \Theta
(q^2-\mu^2)\Delta_R({x\over x^{\prime}},Q_t^2)
{Q_t^2 \over (\mbox{\boldmath $q$}+
\mbox{\boldmath $Q_t$})^2} 
f(x^{\prime},(\mbox{\boldmath $q$}+
\mbox{\boldmath $Q_t$})^2) +O(\mu^2/Q_t^2)
\label{bfklr}
\end{equation} 
where $\Delta_R$ which screens the $1/z$ singularity is given by: 
\begin{equation}
\Delta_R(z,Q_t^2)=z^{\bar \alpha_s ln (Q_t^2/ \mu^2)}= 
exp\left(-\bar \alpha_s\int_z^1{dz^{\prime}
\over z^{\prime}}\int_{\mu^2}^{Q_t^2}{dq^2\over q^2}\right)
\label{deltar}
\end{equation}  
and 
\begin{equation}
\hat f^0(x,Q_t^2)= \int_x^1{dx^{\prime}\over 
x^{\prime}} \Delta_R({x\over x^{\prime}},Q_t^2)
{df^0(x^{\prime},Q^2_t)\over dln(1/x^{\prime})}
\label{hatf0}
\end{equation} 
 Equation (\ref{bfklr}) sums  the ladder diagrams (see Fig. 2)  
with the reggeized gluon exchange along the chain  with the gluon 
trajectory $\alpha_G(Q_t^2) = 1-{\bar \alpha_s\over 2} ln(Q^2_t/\mu^2)$.\\

For the fixed coupling case  eq. (\ref{bfkl}) can be solved 
analytically and the leading behaviour of its solution 
at small $x$ is given by the 
following expression:
\begin{equation} 
f(x,Q_t^2) \sim (Q_t^2)^{{1\over 2}} {x^{-\lambda_{BFKL}}\over 
\sqrt{ln({1\over x})}} exp\left(-{ln^2(Q_t^2/\bar Q^2)\over 2 \lambda^"
ln(1/x)} \right)
\label{bfkls}
\end{equation} 
with 
\begin{equation}
\lambda_{BFKL}=4 ln(2) \bar \alpha_s
\label{pombfkl}
\end{equation} 
\begin{equation} 
\lambda^"=\bar \alpha_s 28 \zeta(3) 
\label{diff}
\end{equation}
where the Riemann zeta function $\zeta(3) \approx 1.202$.  The 
parameter $\bar Q$ is of nonperturbative origin.\\

The quantity $1+ \lambda_{BFKL}$ is equal to the intercept of the so -  
called BFKL pomeron. Its potentially large magnitude ($\sim 1.5$) 
should be contrasted with the intercept $\alpha_{soft} \approx 1.08$ 
of the (effective) "soft" pomeron which has been determined 
from the phenomenological analysis of the high energy behaviour 
of hadronic and photoproduction total cross-sections \cite{DOLA}.\\   
  
The solution of the BFKL equation 
reflects its diffusion pattern which  
is the direct consequence of the absence of transverse momentum ordering 
along the gluon chain.  In this respect the BFKL dynamics is different 
from that based on the (leading order) Altarelli-Parisi evolution which 
corresponds to  strongly -  ordered transverse momenta.  
The interrelation between the diffusion of transverse momenta towards 
both the infrared and ultraviolet regions {\bf and} the increase of gluon 
distributions 
with decreasing $x$ is a  characteristic property of QCD at low $x$.  
It has important consequences for the structure of the hadronic final 
state in deep inelastic scattering at small $x$.\\

In practice one introduces the running coupling $\bar \alpha_s(Q_t^2)$ 
in the BFKL equation (\ref{bfkl}). This requires the introduction of an 
infrared 
cut-off to prevent entering the infrared region where the 
coupling becomes large. The effective intercept $\lambda_{BFKL}$ 
found by numerically solving the equation depends weakly  
on the magnitude of this cut-off \cite{KMS2}. The impact of the momentum 
cut-offs on the solution of the BFKL equation has also been discussed 
in refs. \cite{PVLC,MCDG}.\\

The solution (\ref{bfkls}) of the BFKL equation is obtained most 
directly by solving  the corresponding equation for the moment 
function  
\begin{equation}
\bar f(\omega, Q_t^2)= \int_0^1 {dx\over x} x^{\omega} f(x,q_t^2)
\label{momf}
\end{equation}
$$
\bar f(\omega,Q_t^2)=\bar f^0(\omega,Q_t^2)+
$$
\begin{equation}
{\bar \alpha_s \over \omega}  \int {d^2 q\over \pi q^2}
\left[{Q_t^2 \over (\mbox{\boldmath $q$}+
\mbox{\boldmath $Q_t$})^2} 
\bar f(\omega,(\mbox{\boldmath $q$}+
\mbox{\boldmath $Q_t$})^2)-\bar f(\omega,Q_t^2)\Theta(Q_t^2-q^2)\right]
\label{bfklm}
\end{equation}
This equation can be diagonalised by a Mellin transform.  The solution 
for the Mellin transform $\tilde f(\omega, \gamma)$ of the moment 
function $\bar f(\omega,Q_t^2)$ is: 
\begin{equation}
\tilde f(\omega, \gamma)= {\tilde f^0(\omega, \gamma)\over 
1-{\bar \alpha_s \over \omega} \tilde K(\gamma)}
\label{bfklsg}
\end{equation}
where 
\begin{equation}
\tilde K(\gamma)=2\Psi(1)-\Psi(\gamma)-\Psi(1-\gamma)
\label{kg}
\end{equation}
is the Mellin transform of the kernel of eq. (\ref{bfklm}). The function 
$\Psi(z)$ is the logarithmic derivative of the Euler $\Gamma$ function. 
The Mellin transform $\tilde f(\omega, \gamma)$ is defined by: 
\begin{equation}
\tilde f(\omega, \gamma)=\int_0^{\infty}{dQ_t^2\over Q_t^2}
(Q_t^2)^{-\gamma}\bar f(\omega,Q_t^2), 
\label{mt}
\end{equation}
and hence the function $\bar f(\omega,Q_t^2)$ is related to 
$\tilde f(\omega, \gamma)$ through the inverse Mellin transform 
\begin{equation}
\bar f(\omega, Q_t^2)= {1\over 2 \pi i} \int_{1/2 - i \infty}^{1/2 +i \infty} 
d\gamma (Q_t^2)^{\gamma}\tilde f(\omega, \gamma). 
\label{imt}
\end{equation} 

The poles of $\tilde f(\omega, \gamma)$ in the $\gamma$ plane define the 
anomalous dimensions of the 
moment function $\bar f(\omega, Q_t^2)$ \cite{JAR}.  
The (leading twist) anomalous  
dimension $ \gamma_{gg}(\omega,\bar \alpha_s)$ of 
$\bar f(\omega,Q_t^2)$ gives  the following behaviour of 
$\bar f(\omega,Q_t^2)$ 
at large $Q_t^2$ 
\begin{equation}
\bar f(\omega,Q_t^2) = 
\tilde f^0(\omega,\gamma=\gamma_{gg}(\omega,\bar \alpha_s)) 
\gamma_{gg}(\omega,\bar \alpha_s)
R(\alpha_s,\omega) 
(Q_t^2)^{ \gamma_{gg}(\omega,\bar \alpha_s)}
\label{adbeh}
\end{equation}
where 
\begin{equation}
R(\alpha_s,\omega) =-\left[  {\bar \alpha_s \over \omega} 
\gamma_{gg}(\omega,\bar \alpha_s){d \tilde K(\gamma)\over 
d\gamma} |_{\gamma= \gamma_{gg}(\omega,\bar \alpha_s)}\right]^{-1}. 
\label{rrr}
\end{equation}
The anomalous dimension will also, of course, control 
the large $Q^2$ behaviour of the moment function 
$\bar g(\omega,Q^2)$ of the integrated gluon distribution 
\begin{equation}
 \bar g(\omega,Q^2)=\int_0^{Q^2}{dQ_t^2\over Q_t^2}f(\omega,Q_t^2)
\label{gintm}
\end {equation}
which has the  following form: 
\begin{equation}
\bar g(\omega,Q^2)=R(\alpha_s,\omega) \bar g^0(\omega)\left
({Q^2\over Q_0^2}\right)^{ \gamma_{gg}(\omega,\bar \alpha_s)}
\label{intgm}
\end{equation} 
where we have introduced the moment function of the input distribution 
\begin{equation}
\bar g^0(\omega)=\tilde f^0(\omega,\gamma=\gamma_{gg}(\omega,\bar \alpha_s))
(Q_0^2)^
{ \gamma_{gg}(\omega,\bar \alpha_s)}. 
\label{g0}
\end{equation}
Equations (\ref{intgm})and (\ref{g0}) follow directly from  equations 
(\ref{adbeh}), (\ref{rrr}) and (\ref{gintm}). 
It may be seen from eq. (\ref{intgm}) that the BFKL singularity 
affects  
the "starting" gluon distribution at $Q^2=Q_0^2$ 
through  
the factor $R$ \cite{CIAFKT}.\\  

It follows from eq.(\ref{kg}) that the anomalous dimension $\gamma_{gg}(
\omega,\bar \alpha_s))$  
is the solution of the following equation:
\begin{equation}
{\bar \alpha_s\over \omega} \tilde K( \gamma_{gg}(\omega,\bar \alpha_s))=1. 
\label{adeq}
\end{equation}
It is a function of only one variable ${\bar \alpha_s \over \omega}$ 
i.e. $ \gamma_{gg}(\omega,\bar \alpha_s) \rightarrow 
 \gamma_{gg}({\bar \alpha_s\over \omega})$.  
The solution of  eq. (\ref{adeq}) makes it possible  
to obtain the anomalous dimension 
$ \gamma_{gg}({\bar \alpha_s\over \omega})$
as a power series of ${\bar \alpha_s\over \omega}$ \cite{JAR}
\begin{equation}
 \gamma_{gg}({\bar \alpha_s\over \omega})= \sum_{n=1}^{\infty} 
 c_n \left (
{\bar \alpha_s \over \omega}\right)^n . 
\label{adexp}
\end{equation}
This power series corresponds to the leading $ln(1/z)$ 
expansion of the splitting function $P_{gg}(z,\alpha_s)$ 
\begin{equation}
zP_{gg}(z,\alpha_s)=\bar \alpha_s 
\sum_{n=1}^{\infty} c_n {(\bar \alpha_s ln(1/z))^{n-1}\over (n-1)!}
\label{exppgg}
\end{equation}
which controls the 
evolution of the gluon distribution. 
\\

The structure functions $F_{2,L}(x,Q^2)$ are  described  at small $x$ 
by the 
diagram of Fig.3 which gives the following relation between 
the structure functions and the unintegrated distribution $f$: 
\begin{equation}
F_{2,L}(x,Q^2)=\int_x^1{dx^{\prime}\over x^{\prime}}\int 
{dQ_t^2\over Q_t^2}F^{box}_{2,L}(
x^{\prime},Q_t^2,Q^2)f({x\over x^{\prime}},Q_t^2). 
\label{ktfac}
\end{equation}
The functions  $F^{box}_{2,L}(x^{\prime},Q_t^2,Q^2)$ may be regarded as  the 
structure 
functions of the off-shell gluons with  virtuality  
$Q_t^2$.  
They are described by the quark box (and crossed box) diagram contributions 
to the 
photon-gluon interaction in the upper part of the diagram of Fig. 3.  
The small $x$ behaviour of the structure functions reflects the small 
$z$ ($z = x/x^{\prime}$) behaviour of the gluon distribution $f(z,Q_t^2)$.\\

Equation (\ref{ktfac}) is an example of the "$k_t$ factorization theorem" 
which relates measurable quantities (like DIS structure functions) to 
the convolution in both longitudinal as well as in transverse momenta of the 
universal gluon distribution $f(z,Q_t^2)$ with the cross-section 
(or structure function) describing the interaction of the "off-shell" gluon 
with the hard probe \cite{CIAFKT,KTFAC}.  
The $k_t$ factorization theorem is the basic tool for 
calculating the observable quantities in the small $x$ region in terms of the 
(unintegrated) gluon distribution $f$ which is the solution of the BFKL 
equation.\\ 

The leading - twist part of the $k_t$ factorization formula can be rewritten 
in a collinear factorization form.  The leading small $x$ effects are then 
automatically resummed in the  
 splitting functions and in the coefficient functions. The $k_t$ 
factorization theorem   can in fact be used as the tool for calculating 
these quantities.   Thus, for instance, the moment function $\bar 
P_{qg}(\omega, 
\alpha_s)$ of the splitting function is represented in the following form 
(in the DIS scheme): 

\begin{equation}
\bar P_{qg}(\omega,\alpha_s)= 
{\gamma_{gg}^{2}({\bar \alpha_s\over \omega})\tilde F^{box}_{2}
\left(\omega=0,\gamma= \gamma_{gg}({\bar \alpha_s\over \omega})\right)
\over 2\sum_i e_i^2 }
\label{pqgf}
\end{equation} 
where  $\tilde F^{box}_{2}(
\omega,\gamma)$ is the Mellin transform of the moment function 
$\bar F^{box}_{2}(
\omega, Q_t^2,Q^2)$ i.e. 
\begin{equation}
\bar F^{box}_{2}(
\omega, Q_t^2,Q^2)={1\over 2 \pi i} \int_{1/2-i\infty}^{1/2+i\infty} 
d\gamma \tilde F^{box}_{2}(
\omega,\gamma)\left(Q^2\over Q_t^2 \right)^{\gamma}
\label{mbox}
\end{equation}

Representation (\ref{pqgf}) generates the following  
expansion of the splitting function $P_{qg}(z,\alpha_s)$ at small $z$:  
\begin{equation}
zP_{qg}(z,\alpha_s)={\alpha_s\over 2 \pi}zP^{(0)}(z) +
(\bar \alpha_s)^2 \sum_{n=1}^{\infty}b_n
{[\bar \alpha_s ln(1/z)]^{n-1}\over (n-1)!}
\label{zpqgf}
\end{equation}
  The first term on the right hand side of eq. (\ref{zpqgf})
vanishes at $z=0$.  It should be noted that 
 the  splitting function $P_{qg}$ 
is formally non-leading at small $z$ when compared with the splitting  
function $P_{gg}$ .   
For moderately small values of $z$ however,  
when the first few terms in the expansions (\ref{adexp}) and (\ref{zpqgf})
dominate, the BFKL effects can be much more important 
in $P_{qg}$  than in $P_{gg}$.  
This comes from the fact that in the expansion (\ref{zpqgf}) 
all coefficients $b_n$ are different from zero while in eq. (\ref{adexp}) 
we have $c_2=c_3=0$ \cite{JAR}.  
The small $x$ resummation effects within the conventional QCD evolution 
formalism have recently been discussed in refs. \cite{EKL,HBRW,BFORTE,FRT}. 
In Fig. 4 we show the results of this analysis for the structure function 
$F_2(x,Q^2)$ \cite {HBRW} where the starting 
parton distributions have been assumed to be non-singular.  In this figure we 
compare 
the standard two-loop results (dashed curves) with those which contain 
the leading $ln(1/x)$ resummations.  The dot-dashed curve corresponds 
to keeping the small $x$ corrections only in the gluon anomalous dimension 
while 
the solid and dotted curves show the effect of including the small 
$x$ resummation in the splitting function $P_{qg}$ as well.  (The two curves 
were obtained by applying two different prescriptions of 
implementing the momentum 
sum rule which is violated in the $LL1/x$ approximation).  One can see from 
this figure that at the moderately small values of $x$ which are relevant for 
the HERA measurements,   the small $x$ resummation effects in the 
splitting function $P_{qg}$ have a much stronger impact on $F_{2}$ than 
the small $x$ resummation in the splitting function $P_{gg}$. This  
reflects  the fact, which has already been mentioned 
above, that  in the expansion (\ref{zpqgf}) 
all coefficients $b_n$ are different from zero while in eq. (\ref{adexp}) 
we have $c_2=c_3=0$.\\

A more general treatment of the gluon ladder than that which follows 
from the BFKL formalism is  provided by 
the CCFM equation based on angular ordering along the gluon chain 
\cite{CCFM,KMS1}.  
This equation embodies both the BFKL equation at small $x$ and the 
conventional Altarelli-Parisi evolution at large $x$.  
The unintegrated gluon distribution $f$ now acquires   
dependence upon an additional scale $Q$ 
which specifies the maximal angle of gluon 
emission.  
The CCFM equation has a form analogous to that of the "folded" BFKL equation 
(\ref{bfklr}): 
$$
f(x,Q_t^2,Q^2)=\hat f^0(x,Q_t^2,Q^2)+ $$
\begin{equation}
\bar \alpha_s
\int_x^1{dx^{\prime}\over 
x^{\prime}} \int {d^2 q\over \pi q^2} \Theta
(Q-qx/x^{\prime})\Delta_R({x\over x^{\prime}},Q_t^2,q^2)
{Q_t^2 \over (\mbox{\boldmath $q$}+
\mbox{\boldmath $Q_t$})^2} 
f(x^{\prime},(\mbox{\boldmath $q$}+
\mbox{\boldmath $Q_t$})^2,q^2))      
\label{ccfm}
\end{equation}
where the theta function $\Theta(Q-qx/x^{\prime})$ reflects the angular 
ordering constraint on the emitted gluon.  
The "non-Sudakov" form-factor $\Delta_R
(z,Q_t^2,q^2  )$ is now given by the following formula: 
\begin{equation}
\Delta_R(z,Q_t^2,q^2)=exp\left[-\bar \alpha_s\int_z^1 {dz^{\prime}
\over z^{\prime}} \int {dq^{\prime 2}
\over q^{\prime 2}}\Theta (q^{\prime 2}-(qz^{\prime})^2)
\Theta (Q_t^2-q^{\prime 2})\right]
\label{ns}
\end{equation}
Eq.(\ref{ccfm}) still contains only the singular term of the 
$g \rightarrow gg$ splitting function  
at small $z$. Its generalization which would  
include 
remaining parts of this vertex (as well as quarks) is possible.  
The numerical analysis of this equation was presented in ref. \cite{KMS1}.\\

In Fig. 5 we show the results for the structure function $F_2$ calculated 
from the $k_t$ factorization theorem using the function $f$ obtained from 
the CCFM equation \cite{CCFMF2}.    
We confront these predictions with the most recent data 
from the H1 and ZEUS collaborations at HERA \cite{H1,ZEUS} as well as 
with the results 
of the analysis which was based on the Altarelli-Parisi equation alone 
without the small $x$ resummation effects being included in the formalism 
\cite{MRS,GRV}.   
In the latter case the singular small $x$ behaviour of the gluon 
and sea quark distributions  
 has to be introduced in the parametrization of the starting 
distributions at the moderately large reference scale $Q^2=Q_0^2$   
 (i.e. $Q_0^2 \approx 4 GeV^2$ or so) \cite{MRS}.  One can also 
generate steep behaviour dynamically starting from  
non-singular "valence-like" parton distributions at some very low 
scale $Q_0^2=0.35GeV^2$ \cite{GRV}. In the latter case the gluon and sea 
quark 
distributions exhibit  "double logarithmic behaviour" \cite{DL} 
\begin{equation}
F_2(x,Q^2) \sim exp \left(2\sqrt{\xi(Q^2,Q_0^2)ln(1/x)}\right)
\label{dlog}
\end{equation}
where 
\begin{equation}
\xi(Q^2,Q_0^2)=\int_{Q_0^2}^{Q^2}{dq^2\over q^2}{3\alpha_s(q^2)\over \pi} . 
\label{evlength}
\end{equation} 
For very small values of the scale $Q_0^2$ the evolution length $\xi(Q^2,
Q_0^2)$  
can become large for moderate and large values of $Q^2$ and the "double 
logarithmic" behaviour (\ref{dlog}) is, within the limited region of $x$,  
similar to that corresponding to the power like increase of the type 
$x^{-\lambda}$, $\lambda \approx 0.3$.  This explains similarity between 
the theoretical curves presented in Fig. 5.  The theoretical results 
also show  that an inclusive quantity like $F_2$ is not the 
best discriminator for revealing the dynamical details at low $x$.  
One may however hope that this can be provided by studying the structure 
of the hadronic final state in deep inelastic scattering and this possibility 
will be briefly discussed in  sec. 4.\\

Although the BFKL and CCFM equations for the gluon distribution become 
equivalent in the leading $ln(1/x)$ approximation they begin to differ if 
 less inclusive quantities  are considered. 
In the case of the observables which are not infrared safe 
(like multiplicity of jets etc.) the CCFM formalism generates  double 
logarithmic $ln(1/x)$ terms i.e.  terms which contain powers of 
$\alpha_sln^2(1/x)$.  This is of course closely related to the fact that 
the angular ordering constraint serves also as the infrared regulator 
and so the transverse momentum integrations generate additional powers 
of $ln(1/x)$ besides the "conventional" ones which come from the 
integration over the longitudinal momenta.  Also the virtual corrections 
contain  double logarithmic terms as well.  This can easily be seen from the 
definition of the "non-Sudakov" form factor (\ref{ns}).  The double logarithmic 
terms exactly cancel between the real emission contributions and virtual 
corrections to the gluon distribution but remain present in the less inclusive 
quantities.  On the other hand in the BFKL case  only the single 
logarithmic terms 
(i.e. powers of $\alpha_s ln(1/x)$) are 
present in all quantities but the observables  which are not 
infrared safe diverge  for $\mu^2 \rightarrow 0$ where $\mu$ is the resolution 
parameter.  The gluon distribution itself remains of course 
finite in this limit.\\

Several new interesting results have been obtained in the formal 
studies of the high energy (or small $x$) limit in QCD which go beyond the 
leading logarithmic approximation \cite{LIPATOV,EFACT,EFACT1,NLOR1,NLOR2}. 
The important theoretical tool 
in this case is the effective field theory where the basic objects are 
the   reggeized gluons and the effective action of this effective field 
theory   
obeys conformal invariance \cite{EFACT,EFACT1}. Theoretical analysis 
simplifies in the large $N_c$ limit. One can discuss both the pomeron 
which appears as the bound state of two (reggeized) gluons, and the 
odderon (i.e. the bound state of three reggeized gluons), as well as bound 
states of many reggeized gluons \cite{ODD}. Conformal 
invariance is also very useful 
for analysing the BFKL pomeron away from the forward direction 
\cite{ELSC1,ELSC2} as well as the  the triple pomeron and more complicated 
vertices \cite{CONFTP}.\\

The genuine next-to-leading ln$(1/x)$ corrections to the BFKL 
equation can be present in all relevant 
quantities i.e. in the particle-particle-reggeon vertex, 
the reggeon-reggeon-particle vertex and in the gluon 
Regge trajectory \cite{NLOR1,NLOR2}.  
(The reggeon here corresponds to the reggeized gluon).  Besides that 
one has also to include  additional region of phase-space 
which goes beyond strong ordering of longitudinal momenta. \\ 

It should finally be emphasised that in  impact parameter representation 
the BFKL equation offers an 
interesting  
interpretation in terms of colour dipoles \cite{DIPOLE}. \\

\section*{4. Small $x$ behaviour of the nonsignlet unpolarized and polarized 
structure functions}
The discussion presented in the previous Section concerned the small $x$ 
behaviour of the singlet structure function which was driven by the gluon 
through the $g \rightarrow q \bar q$ transition. 
The gluons of course decouple from the non-singlet channel and the 
mechanism of generating the small $x$ behaviour in this case is different.\\

The simple Regge pole exchange model predicts in this case that 
\begin{equation}
F_2^{NS}(x,Q^2)=F_2^{p}(x,Q^2)-F_2^n(x,Q^2) \sim x^{1-\alpha_{A_2}(0)}
\label{a2}
\end{equation}
where $\alpha_{A_2}(0)$ is the intercept of the $A_2$ Regge 
trajectory.  For $\alpha_{A_2}(0) \approx 1/2$ this behaviour is stable 
against  leading order QCD evolution.  This follows from the fact that  
the leading singularity of the moment $\gamma_{qq}(\omega)$ of the  
splitting function $P_{qq}(z)$: 
\begin{equation}
\gamma(\omega)=\int_0^1 {dz\over z} z^{\omega}P_{qq}(z)
\label{gqq}
\end{equation}
is located at $\omega=0$ and so 
 the (nonperturbative) $A_2$ Regge pole at $\omega=\alpha_{A_2}(0) \approx 
1/2$  remains the leading 
singularity controlling the small $x$ behaviour of the non-singlet 
structure function.\\ 

The novel feature of the non-singlet channel is the appearence of the 
{\bf double} logarithmic  terms i.e. powers of 
$\alpha_s ln^2(1/x)$ at each order of the perturbative 
expansion \cite{GORSHKOV,JK2,KL,EMR,BER}.  
These double logarithmic terms are generated by the 
ladder diagrams with  quark (antiquark) exchange along the chain.   
The ladder diagrams can acquire corrections from the "bremsstrahlung" 
contributions \cite{KL,BER}  
which do not vanish for the polarized structure function 
$g_1^{NS}(x,Q^2)$ \cite{BER}.\\

In the approximation where the leading double logarithmic terms 
are generated by ladder diagrams illustrated in Fig. 6 the 
unintegrated non-singlet quark distribution $f_q^{NS}(x,k^2_t)$ 
($q^{NS}=u+\bar u - d -\bar d$) satisfies 
the following integral equation  :     
\begin{equation}
f_q^{NS}(x,Q_t^2)=f_{q0}^{NS}(x,Q_t^2)+ \tilde \alpha_s 
\int_x^1{dz\over z}\int_{Q_0^2}^{{Q_t^2\over z}} {dQ_t^{\prime 2}\over 
Q_t^{\prime 2}}f_q^{NS}({x\over z},Q_t^{\prime 2})
\label{dleq}
\end{equation}
where 
\begin{equation}
\tilde \alpha_s = {2 \over 3 \pi} \alpha_s 
\label{atil}
\end{equation} 
and $Q_0^2$ is the infrared cut-off parameter. 
The unintegrated distribution $f_q^{NS}(x,Q_t^2)$ is, as usual, related 
in the following way to the scale dependent (nonsinglet) quark distribution 
$q^{NS}(x,Q^2)$: 
\begin{equation}
q^{NS}(x,Q^2)=\int^{Q^2}{dQ_t^2\over Q_t^2}f_q^{NS}(x,Q_t^2). 
\label{intd}
\end{equation}
The upper limit $Q_t^2/z$ in the integral equation (\ref{dleq}) follows 
from the 
requirement that the virtuality of the quark at the end of the chain 
is dominated by $Q_t^2$. A possible non-perturbative $A_2$ reggeon contribution 
has to be introduced in the driving term i.e. 
\begin{equation}
f_{q0}^{NS}(x,Q_t^2) \sim x^{-\alpha_{A_2}(0)}
\label{a2driv}
\end{equation}
at small $x$.\\
  
Equation (\ref{dleq}) implies the following equation 
for the moment function $\bar f_q^{NS}(\omega,Q_t^2)$ 
\begin{equation}
\bar f_q^{NS}(\omega,Q_t^2)=\bar f_{q0}^{NS}(\omega,Q_t^2)+ 
{\tilde \alpha_s \over \omega} \left[\int_{Q_0^2}^{Q_t^2} 
{dQ_t^{\prime 2}\over 
Q_t^{\prime 2}}\bar f_q^{NS}(\omega,Q_t^{\prime 2})+ 
\int_{Q_t^2}^{\infty} {dQ_t^{\prime 2}\over 
Q_t^{\prime 2}}\left({Q_t^2 \over Q_t^{\prime 2}}\right)^{\omega}
\bar f_q^{NS}(\omega,Q_t^{\prime 2})\right]
\label{dleqm}
\end{equation}
Equation (\ref{dleqm}) follows from (\ref{dleq}) after taking  into 
account the following relation: 
\begin{equation}
\int_0^1{dz\over z}z^{\omega}\Theta \left({Q^2_t\over Q^{\prime 2}_t} - z
\right)=
{1\over \omega}\left[\Theta(Q_t^2-Q^{\prime 2}_t)+
\left({Q^2_t\over Q^{\prime 2}_t}\right)^
{\omega}\Theta (Q^{\prime 2}_t-Q^{2}_t)\right]. 
\label{theta}
\end{equation}
For fixed coupling $\tilde \alpha_s$  equation (\ref{dleqm})
 can be solved analytically.  
Assuming for simplicity that the inhomogeneous term is independent 
of $Q_t^2$ (i.e. that $\bar f_{q0}^{NS}(\omega,Q_t^2) = C(\omega)$ )
we get the following solution of  eq.(\ref{dleqm}): 
\begin{equation}
\bar f_q^{NS}(\omega,Q_t^2)=C(\omega)R(\tilde \alpha_s,  \omega) 
\left({Q_t^2\over Q_0^2}\right)^{ \gamma^{-}(\tilde \alpha_s,  \omega)}
\label{solm}
\end{equation}
where 
\begin{equation}
\gamma^{-}(\tilde \alpha_s, \omega) = {\omega - \sqrt{\omega^2 - 4 \tilde 
 \alpha_s}\over 2}
\label{anomd}
\end{equation}
and
\begin{equation}
R(\tilde \alpha_s,  \omega)= {\omega \gamma^{-}(\tilde \alpha_s, \omega)\over 
\tilde \alpha_s}. 
\label{r}
\end{equation}
Equation (\ref{anomd}) defines the anomalous dimension of the 
 moment of the non-singlet quark distribution in which 
 the double logarithmic $ln(1/x)$ terms i.e. the powers of ${\alpha_s \over 
\omega^2}$ have been resummed to all orders.  It can be seen from (\ref{anomd}) 
that this anomalous dimension has a (square root) branch point singularity 
at $\omega=
\bar \omega$ where 
\begin{equation}
\bar \omega= 2 \sqrt{\tilde \alpha_s}. 
\label{barom}
\end{equation} 
This singularity will of course be also present in the moment function $
\bar f_q^{NS}(\omega,Q_t^2)$ itself. It should be noted that in contrast to the 
BFKL singularity whose position above unity was proportional to $\alpha_s$,  
$\bar \omega$ is proportional to $\sqrt{\alpha_s}$ - this being the 
straightforward consequence of the fact that  equation (\ref{dleqm}) 
sums double logarithmic terms $({\alpha_s\over \omega^2})^n$. 
This singularity gives the following contribution to the 
non-singlet quark distribution $f_q^{NS}(x,Q_t^2)$ at small 
$x$:  
\begin{equation} 
f_q^{NS}(x,Q_t^2) \sim {x^{-\bar \omega}\over ln^{3/2}(1/x)}. 
\label{smxns}
\end{equation}
For small values of the QCD  coupling this contribution remains non - leading 
in comparison to the contribution of the $A_2$ Regge pole.\\

 As  has been  mentioned above the 
corresponding integral equation which resums the double logarithmic 
terms  in the spin dependent quark distributions is more 
complicated than the simple ladder equation (\ref{dleq}) 
due to non-vanishing contributions coming from bremsstrahlung diagrams.       
It may however be shown that , at least as far as the non-singlet 
structure function is concerned, these contributions give only a relatively small 
correction to $\bar \omega$.  
The main interest in applying the QCD evolution equations to study  
the spin structure function is that the naive Regge pole expectations based 
on the exchange of  low-lying Regge trajectories become unstable against 
the QCD perturbative "corrections".  The relevant reggeon which contributes 
to $g_1^{NS}(x,Q^2)$ is the  $A_1$  exchange which 
is expected to have a very low intercept $\alpha_{A_1}(0) \le 0$.  
The perturbative singularity generated by the double logarithmic 
$ln(1/x)$ resummation can therefore become much more important 
than in the case of the unpolarized case where it is hidden behind  
  leading $A_2$ exchange contribution.  Even if we restrict ourselves 
to  leading order QCD evolution  
\cite{GGR,BFR}  
then the non-singular $x^{-\alpha_{A_1}(0)}$ behaviour 
(with $\alpha_{A_1}(0) \le 0$ ) becomes unstable as well 
and the polarized quark 
densities acquire singular behaviour:   
\begin{equation}
\Delta q^{NS}(x,Q^2) \sim exp(2 \sqrt{\xi^{NS}(Q^2)ln(1/x)})
\label{dqdl}
\end{equation}
where 
\begin{equation}
\xi^{NS}(Q^2)=\int^{Q^2} {dq^2 \over q^2} \tilde \alpha_s(q^2)
\label{xins}
\end{equation}
This follows from the fact that $\Delta P_{qq}(z)=P_{qq}(z)$ 
where $P_{qq}(z)$  and  $\Delta P_{qq}(z)$ are the splitting functions 
describing the evolution of the spin independent and spin dependent quark 
distributions respectively and from the fact that $P_{qq}(z) \rightarrow 
const$ as $z \rightarrow 0$.\\  

The introduction of the running coupling effects in  equation (\ref{dleqm})
turns the branch point singularity into the series of poles which accumulate 
at $\omega=0$ \cite{JK2}.  The numerical analysis of the corresponding 
integral equation,  
with the running coupling effects taken into account,  
gives an effective slope ,  
\begin{equation}
\lambda(x,Q_t^2)={dln \Delta f_q^{NS}(x,Q_t^2)\over d ln(1/x)}
\label{slope}
\end{equation}
with magnitude $\lambda(x,Q_t^2) \approx 0.2 - 0.3$ at small $x$ 
\cite{JK3}.   
The result of this estimate suggest that a reasonable extrapolation 
of the (non-singlet) polarized quark densities would be to assume an  
$x^{-\lambda}$ behaviour with $\lambda \approx 0.2 - 0.3$.   Similar 
 extrapolations of the spin-dependent quark 
distributions towards the small $x$ region have  
 been assumed in  several recent parametrizations of parton densities 
\cite{BS,GRVOG,GS,BV}. The small $x$ behaviour of the spin dependent 
structure function $g_1$ has also been disscussed in refs. \cite{BASLO,RGRF}. 
 \\

\section*{5. The structure of the hadronic final state in deep inelastic 
scattering at low $x$}
It is expected that absence of transverse momentum ordering along the gluon 
chain,  which leads to the correlation between the increase of the 
structure function  with  decreasing $x$ and the diffusion 
of transverse momentum should reflect itself in the behaviour of 
less inclusive 
quantities than the structure function $F_2(x,Q^2)$.  The dedicated 
measurements of  low $x$ physics which are particularly sensitive 
to this correlation are the deep inelastic plus 
jet events, transverse energy flow 
in deep inelastic scattering, production of jets separated by the large 
rapidity gap and dijet production in deep inelastic scattering.  
The diagrammatic illustration of these measurements is presented in 
Fig. 7.\\

In principle  deep inelastic lepton scattering containing a measured jet 
can provide a very clear test of the BFKL dynamics at low $x$
\cite{MJET,BJET,KMSJET}.  
The idea is to study deep inelastic ($x,Q^2$) events which 
contain an identified jet ($x_j,k_{Tj}^2$) where 
$x<<x_j$ and $Q^2 \approx k_{Tj}^2$.  Since we choose events with 
$Q^2 \approx k_{Tj}^2$ the leading order QCD evolution  (from $k_{Tj}^2$ to 
$Q^2$) is neutralized and attention is focussed on the small $x$, or rather 
small $x/x_j$ behaviour.  The small $x/x_j$ behaviour
of jet production is generated  by the gluon radiation as shown 
in the diagram of Fig. 7a.  Choosing the 
configuration $Q^2 \approx k_{Tj}^2$ we eliminate by definition  
gluon emission which corresponds to strongly ordered transverse momenta 
i.e.  that emission which is responsible for the LO QCD evolution.  
The measurement of jet production in this configuration may therefore test 
more directly the $(x/x_j)^{-\lambda}$ behaviour which is generated 
by the BFKL equation where the transverse momenta are not ordered.  
The recent H1 results concerning  deep inelastic plus jest events 
are consistent with the increase of the cross-section with decreasing 
$x$ as predicted by the BFKL dynamics {\cite{H1JET}. \\

A conceptually similar process is that of the two-jet production, 
with the jets  
separated by a large rapidity gap $\Delta y$,  in hadronic 
collisions or in photoproduction as illustrated in Fig. 7c
\cite{DDUCA,JAMESJ}.  
Besides the characteristic $exp(\lambda \Delta y)$ dependence 
of the two-jet cross-section one expects significant 
weakening of the azimuthal back-to-back correlations 
of the two jets.  This is the direct consequence of the 
absence of transverse momentum ordering along the gluon  
chain in the diagram of Fig. 7c.\\

Another measurement which should be sensitive to the QCD 
dynamics at small $x$ is that of the transverse energy flow in deep inelastic 
lepton scattering in the central region away from the 
current jet and from the proton remnant as illustrated in Fig. 7b. 
\cite{KMSET}.  
The  BFKL dynamics predicts 
in this case a substantial amount of transverse energy  which should 
increase with decreasing $x$.  The experimental data 
are consistent with this theoretical expectation \cite{H1JET}.      
Absence of transverse momentum ordering  also implies weakening of the 
back-to-back azimuthal correlation of dijets produced close 
to the photon fragmentation region (see Fig. 7d) \cite{DICKJET,AGKM}. \\

\section*{6. Summary and conclusions}

In this talk we have briefly described the QCD expectations 
for deep inelastic lepton scattering at low $x$ which  follow from the 
BFKL dynamics.  It leads to the indefinite increase of gluon distributions 
with decreasing $x$ which is correlated with the diffusion of 
 transverse momenta. This increase of gluon distribution 
implies a similar increase of the structure functions through   
$g \rightarrow q \bar q$ transitions.  
Besides discussing the theoretical and 
phenomenological issues related to the description of the structure 
function $F_2$ at low $x$ we have also emphasised 
the role of studying the hadronic final state in deep inelastic scattering.  
\\

The indefinite growth of parton distributions cannot go on forever 
and has to be eventually stopped by parton screening which leads 
to the parton saturation.  Most probably, however,  
the saturation limit is still irrelevant for the small $x$ region 
which is now being probed at HERA. 
\\ 

We have also discussed the small $x$ behaviour 
of the spin dependent structure 
function $g_1$ focussing  for simplicity on its  
non-singlet component which, at small $x$ , can be (approximately) 
described by  ladder diagrams with the quark (antiquark) exchange.  
The novel feature in this case is the appearance of  double logarithmic  
terms i.e.  contributions which contain powers of $\alpha_s ln^2(1/x)$. 
The perturbative QCD effects become significantly amplified for the 
singlet spin structure function due to  mixing with the gluons.  
The simple ladder equation may not however be  applicable 
for an accurate description of the double logarithmic terms in 
the polarized gluon distribution $\Delta G$.\\

We have limited ourselves to the large $Q^2$ region where perturbative 
QCD is expected  to be applicable.  Specific problems of the low $
Q^2$, low $x$ region are discussed in ref. \cite{BBJK}.   
Finally let us point out that the change of the dynamics with 
the relevant scale is clearly visible in the data (see Fig. 1) and its 
satisfactory explanation is perhaps one of the most challenging 
problems of to-day.               
\medskip\medskip\medskip                
     
\section*{Acknowledgments}
I thank Robin Devenish and Mike Whalley for organizing an excellent meeting.  
I thank Barbara Bade\l{}ek, Krzysztof Golec-Biernat, 
Alan Martin and Peter Sutton for most enjoyable research collaborations  
on the problems presented in this lecture.   
I am grateful to Grey College and to Physics 
Department of the University of Durham for their warm hospitality.  
This research has been supported in part by 
 the Polish State Committee for Scientific Research grant 2 P03B 231 08  and 
the EU under contracts n0. CHRX-CT92-0004/CT93-357.\\

{\Large {\bf Figure captions}}
\begin{enumerate}
\item
The total virtual photon cross-section plotted as a function of $W^2$ 
for different values of the photon virtuality $Q^2$.  The curves 
correspond to the theoretical parametrizations \cite{GRV,DOLAF2}. 
In particular the skewed curve is the GRV prediction for 
$Q^2=0.35 GeV^2$.    
The figure is taken from ref. \cite{ALEVY}. 
\item
Diagrammatic representation of the BFKL equation (\ref{bfklr}). 
\item 
Diagrammatic representation of the $k_t$ factorization formula (\ref{ktfac}). 
\item
Theoretical predictions for the structure function $F_2(x,Q^2)$ 
based on the Altarelli-Parisi evolution equations with the leading 
$ln(1/x)$ terms  resummed in both the $P_{gg}$ and $P_{qg}$ splitting 
functions (solid and dotted curves where the dotted curve is based 
on the different prescription for imposing the momentum sum rule). 
The dot-dashed curves show predictions when the leading $ln(1/x)$ 
resummation is included only in the splitting function $P_{gg}$.  
The dashed curves correspond to the two-loop prediction. 
The theoretical curves are confronted with the 1993 data from HERA. 
The figure is taken from ref. \cite{HBRW}. 
\item
A comparison of the HERA measurements of $F_2$ \cite{H1,ZEUS} with 
the predictions based on the $k_t$ factorization formula (\ref{ktfac}) 
using for the unintegrated gluon distributions $f$ the solutions of the 
CCFM equation (\ref{ccfm}) (continuous curve) and of the  
approximate form of this 
equation corresponding to setting 
$\Theta(Q-q)$ in place of $\Theta(Q-qx/x^{\prime})$ and  
$\Delta_R=1$ (dotted curve).  
 We also show the values of $F_2$ obtained from 
collinear factorization using  MRS(A$^{\prime})$ \cite{MRS} and  
GRV \cite{GRV} partons (the figure is taken from  ref. \cite{CCFMF2}). 
\item 
The ladder diagram with quark (antiquark) exchange. 
\item
Diagrammatic representation of the processes testing  BFKL 
dynamics. (a) Deep inelastic scattering with the forward jet. 
(b) $E_T$ flow in deep inelastic scattering. (c) Production 
of jets separated by the large rapidity gap $\Delta y$. (d) 
Dijet production in deep inelastic scattering (the figure is taken from 
ref. \cite{ADMBLOIS}).
\end{enumerate}       

\begin{thebibliography}{9999}
\bibitem{GLR} L.N. Gribov, E.M. Levin and M.G. Ryskin, 
Phys. Rep. {\bf 100} (1983) 1.
\bibitem{BCKK} B. Bade\l{}ek et al., Rev. Mod. Phys. {\bf 64} (1992) 927. 
\bibitem{ADM1} A.D. Martin, Acta. Phys. Polon, {\bf B25} (1994) 265. 
\bibitem{JK1} J. Kwieci\'nski,  Nucl. Phys. B (Proc. Suppl.) {\bf 39 B,C} 
(1995) 58. 
\bibitem{BFKL}E.A. Kuraev, L.N.Lipatov and V.S. Fadin, Zh. Eksp. Teor. Fiz. 
{\bf 72} (1977) 373 (Sov. Phys. JETP {\bf 45} (1977) 199); 
Ya. Ya. Balitzkij and L.N. Lipatov, Yad. Fiz. {\bf 28} (1978) 1597 (Sov. J. 
Nucl. Phys. {\bf 28} (1978) 822);  
J.B. Bronzan and R.L. Sugar, Phys. Rev. {\bf D17} (1978) 585; 
T. Jaroszewicz, Acta. Phys. Polon. {\bf B11} 
(1980) 965. 
\bibitem{LIPATOV}L.N. Lipatov, in "Perturbative QCD", edited 
by A.H. Mueller, (World Scientific, Singapore, 1989), p. 441.
\bibitem{AKMS} A.J. Askew et al., Phys. Rev. {\bf D47} (1993) 3775; 
Phys. Rev. {\bf D49} (1994) 4402. 
\bibitem{H1}H1 collaboration: A. de Roeck et al., preliminary measurements 
to be published in the Proc. of the Workshop on DIS and QCD, Paris, 1995, 
DESY preprint 95 - 152. 
\bibitem{ZEUS}ZEUS collaboration: B. Foster, to be published in the Proc. 
of the Workshop on DIS and QCD, Paris, 1995, DESY preprint 95-193. 
\bibitem{DOLA} A. Donnachie and P.V. Landshoff, Phys. Lett. {\bf B296} (1992) 
257.
\bibitem{PC}P.D.B. Collins, "An Introduction to Regge Theory and High Energy 
Physics", Cambridge University Press, Cambridge, 1977.
\bibitem{IOFFE} B.L. Ioffe, V.A. Khoze and L.N. Lipatov, "Hard Processes", 
North Holland, Amsterdam-Oxford-NewYork-Tokyo, 1984. 
\bibitem{EKARL}J. Ellis and M. Karliner, Phys. Lett. {\bf B231} (1988) 72.
\bibitem{DOLAF2} A. Donnachie and P.V. Landshoff, Z. Phys. {\bf C61} 
(1994) 161.   
\bibitem{ALEVY} A. Levy, these proceedings.  
\bibitem{CIAF}M. Ciafaloni, Nucl. Phys. {\bf B296} (1988) 49. 
\bibitem{CCFM}S. Catani, F. Fiorani and G. Marchesini, 
Phys. Lett. {\bf B234} (1990) 339; Nucl. Phys. {\bf B336} 
(1990) 18; G. Marchesini, in Proceedings of the Workshop "QCD at 200 TeV", 
Erice, Italy, 1990, edited by. L. Cifarelli and Yu. L. Dokshitzer 
(Plenum Press, New York, 1992), p. 183; G. Marchesini, Nucl. Phys.
{\bf B445} (1995) 49. 
\bibitem{KMS1}J. Kwieci\'nski, A.D. Martin, P.J. Sutton, Phys. Rev. 
{\bf D52} (1995) 1445.
\bibitem{BO}B. Andersson, G. Gustafson and J. Samuelsson, Lund preprint 
LU TP 95-13. 
\bibitem{MTK}Yu.L. Dokshitzer et al., Rev. Mod. Phys. {\bf 60} (1988) 373.
\bibitem{KTFAC}S. Catani, M. Ciafaloni and F. Hautmann, Phys. Lett. 
{\bf B242} (1990) 97; Nucl. Phys. {\bf B366} (1991) 657; 
J.C. Collins and R.K. Ellis, Nucl. Phys. {\bf B 360} (1991) 3; 
S. Catani and F. Hautmann, Nucl. Phys. {\bf B427} (1994) 475.
\bibitem{CIAFKT}M. Ciafaloni, Phys. Lett. {\bf 356} (1995) 74.
\bibitem{MK}J. Kwieci\'nski, A.D. Martin, Phys. Lett. {\bf B353} (1995) 
123.    
\bibitem{JAR} T. Jaroszewicz, Phys. Lett. {\bf B116} (1982) 291.
\bibitem{KMS2} A.D. Martin, 
J. Kwieci\'nski  and P.J. Sutton, Nucl. Phys. B (Proc. Suppl.) 
{\bf A29} (1992) 67. 
\bibitem{PVLC} J.C. Collins, P.V. Landshoff, Phys. Lett. {\bf B276} 
(1992) 196. 
\bibitem{MCDG} J.F. McDermott, J.R. Forshaw and G.G. Ross, Phys. Lett. 
{\bf B349} (1995) 189. 
\bibitem{EKL} R.K. Ellis, Z. Kunszt and E.M. Levin, Nucl. Phys. 
{\bf B420} (1994) 517; Erratum-ibid. {\bf B433} (1995) 498. 
\bibitem{HBRW} R.K. Ellis, F. Hautmann and B.R. Webber, 
Phys. Lett. {\bf B348} (1995) 582. 
\bibitem{BFORTE}R.D. Ball and S. Forte, Phys. Lett. {\bf 351} 
(1995) 313.   
\bibitem{FRT} J.R. Forshaw, R.G. Roberts and R.S. Thorne, 
Phys. Lett. {\bf B356} (1995) 79.
\bibitem{CCFMF2}J. Kwieci\'nski, A.D.Martin and P.J. Sutton, 
Durham preprint DTP/95/94. 
\bibitem{MRS}A.D. Martin, R.G. Roberts and W.J. Stirling, 
Phys. Rev. {\bf D50} (1994) 6734; Phys. Lett. {\bf 354} (1995) 155. 
\bibitem{GRV}M. Gl\"uck, E. Reya and A. Vogt, Z.Phys. {\bf C67} (1995) 433. 
\bibitem{DL}A. de Rujula et al., Phys. Rev. {\bf D10} (1974) 1649.  
\bibitem{EFACT}L.N. Lipatov, Nucl. Phys. {\bf B452} (1995) 969; 
talk given at the International 
Conference on Elastic and Diffractive Scattering, "Frontiers 
in Strong Interactions", Chateau de Blois, France, June 1995.  
\bibitem{EFACT1}R. Kirschner, L.N. Lipatov and L. Szymanowski, 
Nucl. Phys. {\bf B425} (1994) 579; Phys. Rev. {\bf D51} (1995) 838. 
\bibitem{ODD} L.D. Fadeev and G.P. Korchemsky, 
Phys. Lett. {\bf B342} (1995) 311; 
G.P. Korchemsky, to be published in the Proc. of the Workshop on DIS and QCD; 
G.P. Korchemsky, Stony Brook preprint ITP-SB-95-25; 
R. Janik, Cracow preprint TPJU-18-95. 
\bibitem{ELSC1}A.H. Mueller and W.K. Tang, Phys. Lett. {\bf B284} 
(1992). 
\bibitem{ELSC2}J. Bartels et al., Phys. Lett. {\bf B348} (1995) 589. 
\bibitem{CONFTP}J. Bartels, M. W\"usthoff, Z. Phys. {\bf C66} (1995) 157; 
J.. Bartels, L.N. Lipatov and M. W\"usthoff, DESY preprint 95-171. 
\bibitem{DIPOLE}A.H. Mueller, Nucl. Phys., {\bf B415} (1994) 373; 
A.H. Mueller and B. Patel, Nucl. Phys. {\bf B425} (1994) 471; 
A.H. Mueller, Nucl. Phys. {\bf B437} (1995) 107; 
Chen Zhang and A.H. Mueller, Nucl. Phys. {\bf B451} (1995) 579. 
\bibitem{NLOR1} V.S.  Fadin, talk given at the International 
Conference on Elastic and Diffractive Scattering, "Frontiers 
in Strong Interactions", Chateau de Blois, France, June 1995. 
\bibitem{NLOR2} A.R. White, talk given at the International 
Conference on Elastic and Diffractive Scattering, "Frontiers 
in Strong Interactions", Chateau de Blois, France, June 1995.   
\bibitem{GORSHKOV} V.G. Gorshkov et al., Yad. Fiz. {\bf 6} (1967) 129 
(Sov. J. Nucl. Phys. {\bf 6} (1968) 95); L.N. Lipatov, 
Zh. Eksp. Teor. Fiz. {\bf 54} (1968) 1520 
(Sov. Phys. JETP {\bf 27} (1968) 814). 
\bibitem{JK2} J. Kwieci\'nski, Phys. Rev. {\bf D26} (1982) 3293. 
\bibitem{KL} R. Kirschner and L.N. Lipatov, Nucl. Phys. {\bf B213} 
(1983) 122. 
\bibitem{EMR} B.I. Ermolaev, S.I. Manayenkov and M.G. Ryskin, DESY preprint 
95-017. 
\bibitem{BER} J. Bartels, B.I. Ermolaev and M.G. Ryskin, DESY preprint 
95 - 124. 
\bibitem{GGR}M.A.Ahmed and G.G. Ross, Phys. Lett. {\bf B56} (1975) 385; 
Nucl. Phys. {\bf B11} (1976) 298; 
G. Altarelli and G. Parisi, Nucl. Phys. {\bf B126} (1977) 298. 
\bibitem{BFR}R.D. Ball, S. Forte and G. Ridolfi, Nucl. Phys. {\bf B444} 
(1995) 287. 
\bibitem{JK3}J. Kwieci\'nski, Durham preprint DTP 95/98 (to appear in Acta 
Physica Polonica). 
\bibitem{BS} C. Bourrely and J. Soffer, Phys. Rev. {\bf D51} (1995) 2108; 
Nucl. Phys. {\bf B445} (1995) 341; Marseille preprint CPT-95/P.3224. 
\bibitem{GRVOG}M. Gl\"uck, E. Reya and W. Vogelsang, Phys. Lett. {\bf B359} 
(1995) 201.
\bibitem{GS} T. Gehrmann and W.J. Stirling, Durham preprint DTP/95/62. 
\bibitem{BV}J. Bl\"umlein and A. Vogt, DESY preprint 95-175. 
\bibitem{BASLO} S.D. Bass, P.V. Landshoff, Phys. Lett. {\bf B336} (1994) 
537.  
\bibitem{RGRF} F.E. Close, R.G. Roberts, Phys. Lett. {\bf B336} (1994) 257.     
\bibitem{ADMBLOIS} A.D. Martin, review talk presented at the Blois 
Workshop "The Heart of the Matter", June 1994, Blois, France. 
\bibitem{MJET}A.H. Mueller, J. Phys. {\bf G17} (1991) 1443. 
\bibitem{BJET}J. Bartels, M. Loewe and A. DeRoeck, Z. Phys. {\bf C54} (1992) 
635.
\bibitem{WKJET}W.K.Tang, Phys. Lett. {\bf B278} (1992) 363. 
\bibitem{KMSJET} J. Kwieci\'nski, A.D. Martin and P.J. Sutton, 
Phys. Rev. {\bf D46} (1992) 921; Phys. Lett. {\bf B287} (1992) 254. 
\bibitem{H1JET} H1 Collaboration, DESY preprint 95 - 108.
\bibitem{DDUCA} V. del Duca, Phys. Rev. {\bf D49} (1994) 4510. 
\bibitem{JAMESJ}W.J. Stirling,  Nucl. Phys. {\bf B423} (1994) 56. 
\bibitem{KMSET} J. Kwieci\'nski, A.D. Martin, P.J. Sutton 
and K.Golec-Biernat, Phys. Rev. {\bf D50} (1994) 217; 
K.Golec-Biernat, J. Kwieci\'nski, A.D. Martin and  P.J. Sutton, 
Phys. Lett. {\bf B335} (1994) 220.   
\bibitem{DICKJET}J.R. Forshaw and R.G. Roberts, RAL preprint 
94-0228. 
\bibitem{AGKM} A.J. Askew et al., Phys. Lett. {\bf B338} (1994) 92. 
\bibitem{BBJK} B. Bade\l{}ek and J. Kwieci\'nski, 
Warsaw preprint IFD/1/94 (to appear in Rev. Mod. Phys.). 
\end{thebibliography}
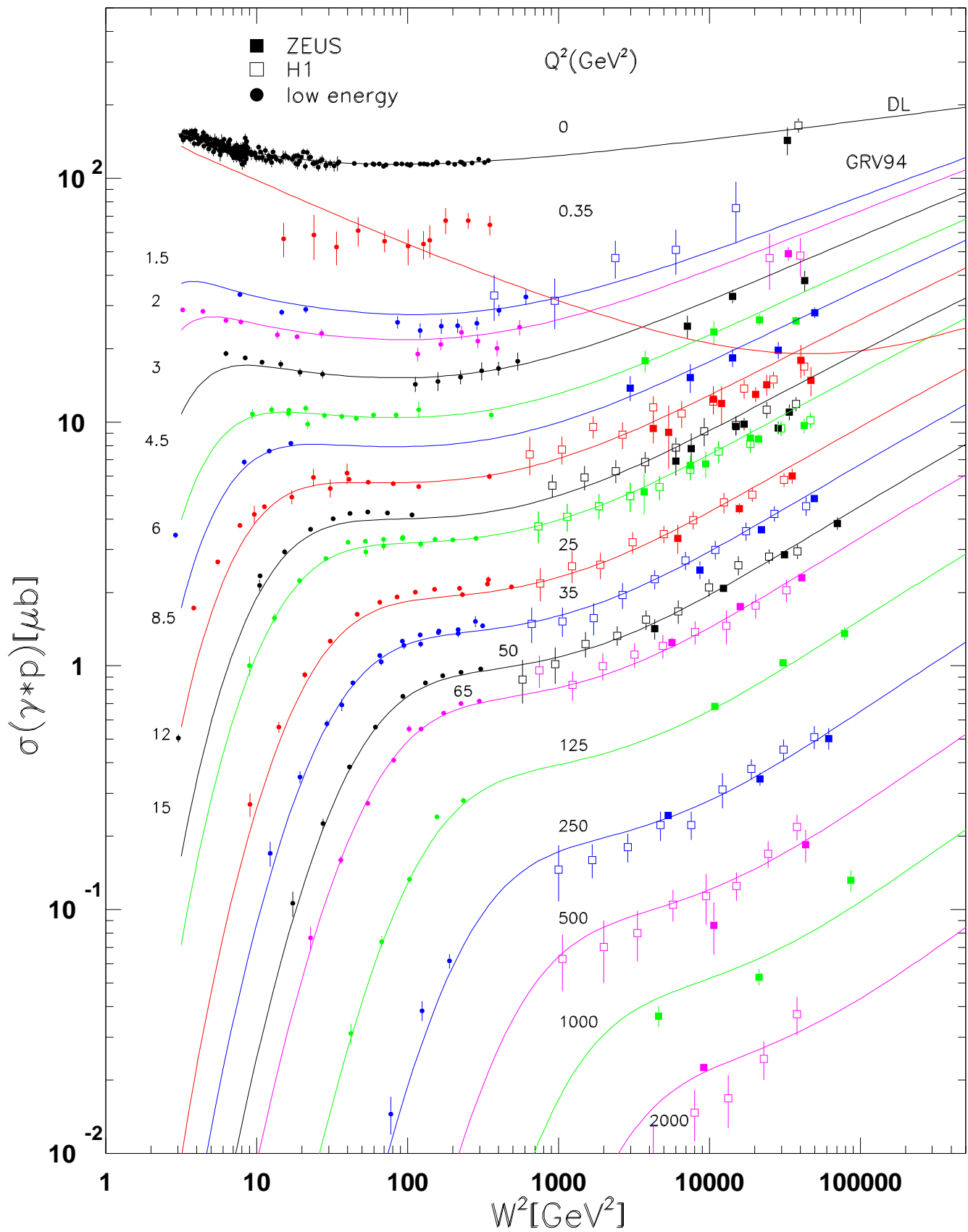
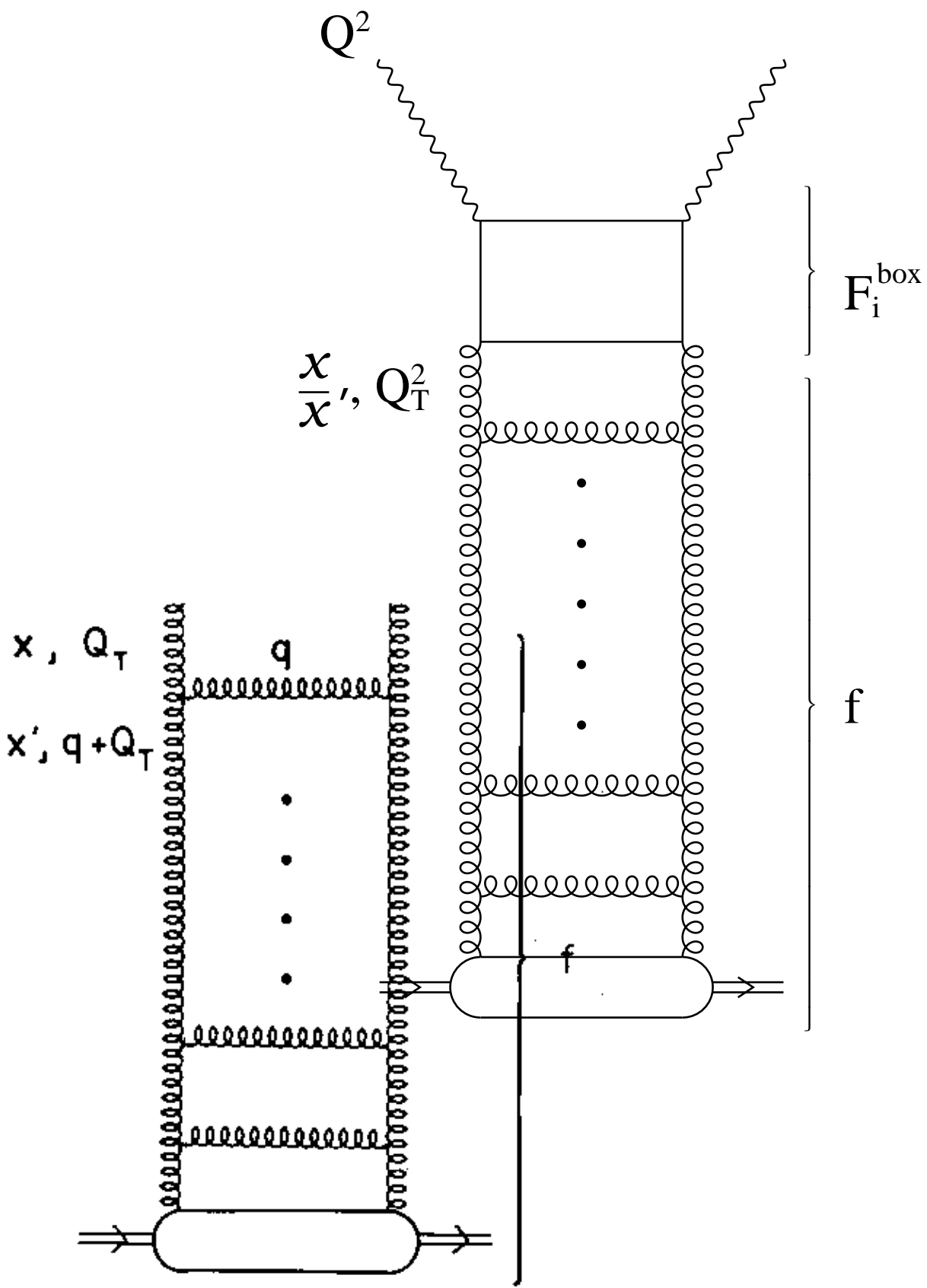
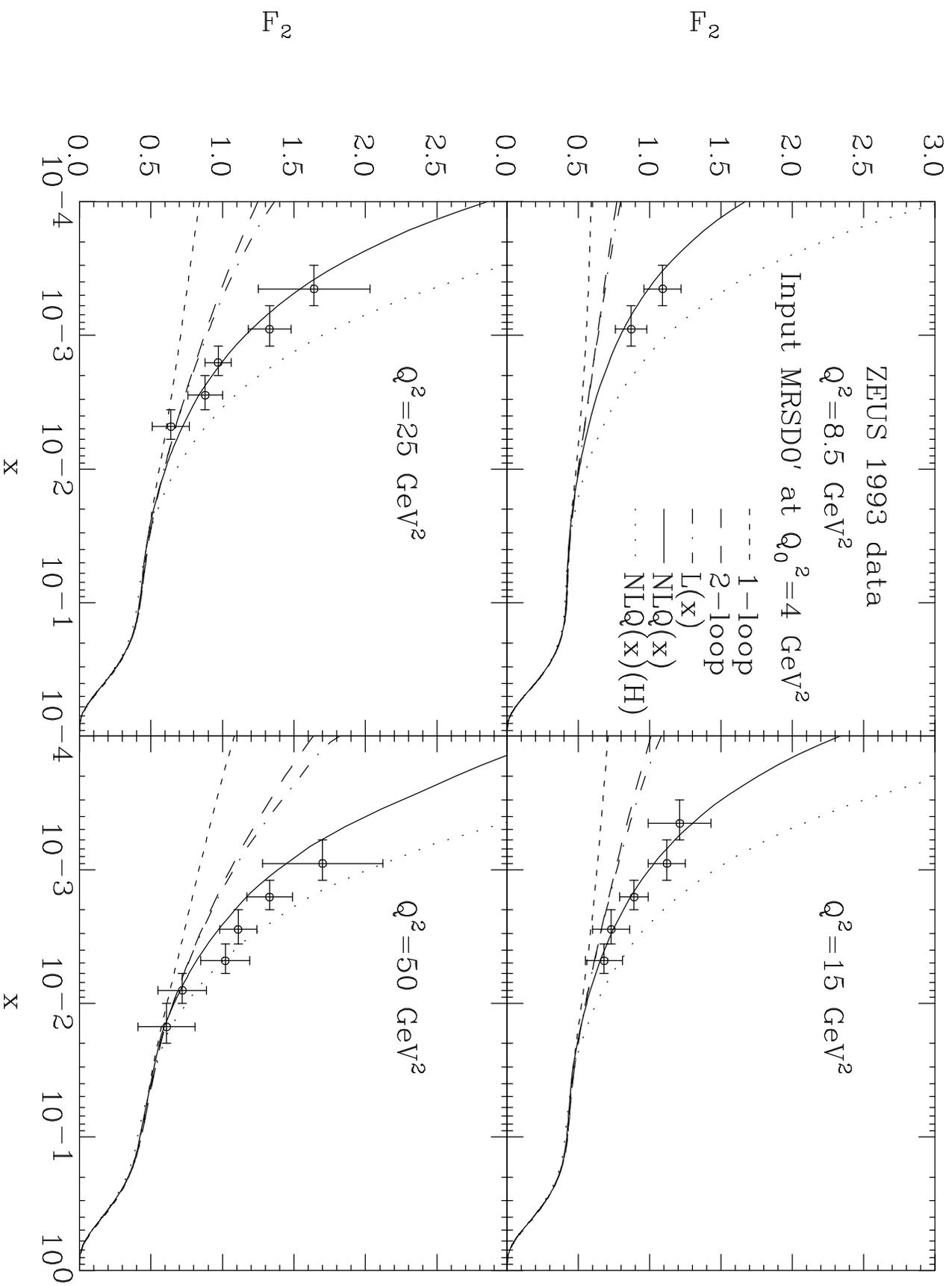
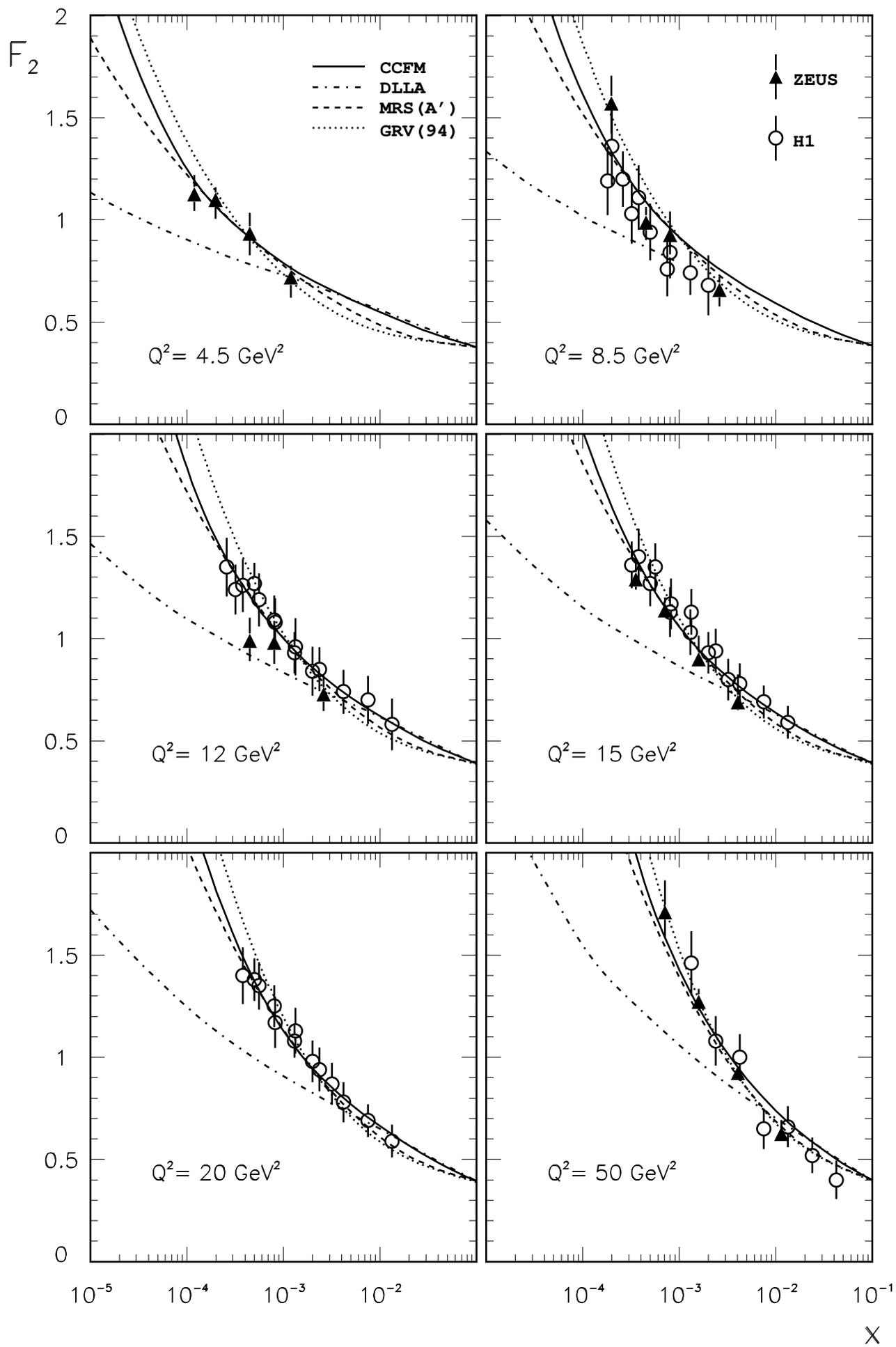
\end{document}